\documentclass[showpacs,preprintnumbers,amsmath,amssymb]{revtex4}
\usepackage{dcolumn}
\usepackage{bm}

\newcommand{\ket}[1]{|#1\rangle}

\begin{document}

\title{Disentanglement and decoherence in two-spin and three-spin
systems under dephasing\\}

\author{Kevin Ann}
\email{kevinann@bu.edu}
\affiliation{Department of Physics, Boston University, Boston, MA 02215}

\author{Gregg Jaeger}
\email{jaeger@bu.edu} \affiliation{Quantum Imaging Lab, Department
of Electrical and Computer Engineering, and Division of Natural
Sciences, Boston University, Boston, MA 02215\\}

\date{\today}

\begin{abstract}
We compare disentanglement and decoherence rates within two-spin and
three-spin entangled systems subjected to all possible combinations
of local and collective pure dephasing noise combinations. In all
cases, the bipartite entanglement decay rate is found to be greater
than or equal to the dephasing-decoherence rates and often
significantly greater. This sharpens previous results for two-spin
systems [Phys. Rev. B 56, 165322 (2003)] and extends them to the
three-spin context.
\end{abstract}


\maketitle


\section{\label{introduction1}INTRODUCTION}
One goal of nanoscience is to understand the behavior of quantum
entanglement in states of multiparticle systems, such as spin
arrays. In such contexts ({\it e.g.} \cite{Wang}), spins are often
called qubits, as we will refer to them here. Entanglement can be
viewed as a resource that allows uniquely quantum phenomena, such as
quantum state teleportation \cite{Laflamme} using quantum memories,
to be exhibited \cite{NC00,JQ06}. Entanglement can be affected by a
number of noise influences, including both vacuum noise \cite{YE04}
and classical noise \cite{YE06}. One important effect is that of the
loss of entanglement, which arises in both these cases. Recently,
for example, it has been noted that in initially entangled two-qubit
systems a sudden loss of entanglement can occur, an effect known as
``entanglement sudden death'' \cite{YE04,YE06}. It is also
interesting to note that in certain situations the irreversible
process of spontaneous emission may lead to the revival of
entanglement \cite{FT06}. An experimental method for the probing of
disentanglement in cavity QED has also been introduced
\cite{SMDZ06}. Recent analyses have analytically examined the
relationship between dephasing decoherence and the reduction of
bipartite entanglement in two-qubit systems under such noise
\cite{YE02,YE03,YE04,TP05,STP06}, as well as in pairs of three-level
systems, {\it i.e.}, two-qutrit systems \cite{DJ06}. Here, we extend
the analytic consideration of this relationship to the case of
three-qubit systems.

Decoherence results from unwanted interactions of a quantum system
with its environment. When such decoherence occurs, it is exhibited
by the decay of off-diagonal elements of the density matrix
describing the system \cite{Omnes}. This may be due to either local
or collective interactions, acting either at the scale of subsystems
or of the composite system; that is, environmental noise may act on
some or all system constituents similarly or differently and on some
not at all \cite{YE03,YE02}. Examples of environments capable of
inducing such interactions include electromagnetic fields and
thermal baths \cite{YE03,YE02,Wang}.  Here, we consider quantum
state decoherence in systems of two or three qubits due to
pure-dephasing noise. In particular, we address the relationship
between degree of entanglement (concurrence) and state coherence
under environmental noise by determining characteristic decay
timescales at the single-qubit, double-qubit, and triple-qubit
scales. We find decoherence timescales to bound corresponding
disentanglement timescales from above in all cases.

We begin by reviewing and sharpening the analysis begun in an
earlier study of two-qubit systems, and then extend that approach to
the case of three-qubit systems, in which more complex combinations
of dephasing noise are relevant. This paper is accordingly organized
as follows. In Sec. II, we revisit the analysis of two-qubit systems
begun by Yu and Eberly in \cite{YE03} that explicitly treats
disentanglement in the operator-sum formalism under local
dephasing-noise environments and explicitly address disentanglement
under collective dephasing noise, and sharpen the conclusion of the
analysis by showing that decoherence times bound disentanglement
times from above in all cases. In Sec. III, we extend the analysis
of the effects of dephasing on coherence and bipartite entanglement
to three-qubit systems in the more complex dephasing noise
environments available there. We conclude with a summary of results
in Section IV.


\section{\label{two-QubitSystem}TWO-QUBIT SYSTEM}

Before considering the effects of dephasing on three-qubit systems,
we complete a previous analysis of two-qubit systems by Yu and
Eberly \cite{YE03}. There, a number of dephasing scenarios involving
characteristic classes of qubit-pair states were discussed; in one
of these, the situation in which collective dephasing, that is,
dephasing between two-qubit-basis states occurs under a non-zero
global magnetic field alone, disentanglement was not explicitly
treated. Here, we discuss disentanglement in this case. The general
conclusion in \cite{YE03} was that bipartite disentanglement can be
much more rapid than local decoherence under local dephasing. Here,
we demonstrate a more general and precise result that accords with
that conclusion: the decoherence timescale bounds the
disentanglement timescale from above for both local and collective
phase-damping noise environments for the two classes of states
introduced there.


\subsection{\label{two-QubitModel}MODEL AND MEASURES}

There exist many related two-component few-state models
corresponding to physical systems the coherence of which may be
influenced.  For example, \cite{YE03} modeled two-qubit systems such
as pairs of atoms in a dephasing-noise environment, whereas
\cite{DJ06} modeled two-qutrit systems corresponding to two
three-level atoms whose two uncoupled excited levels undergo
spontaneous emission to the ground state. In this section, we
examine the model first introduced in \cite{YE03}, namely, that of a
two-qubit system subject to classical dephasing noise.
\newline
Consider a two-qubit system coupled to external sources of noise
that can act both on single qubits individually and on the joint
two-qubit system as a whole. There are many physical situations that
may be described in this way, for example, two spin-1/2 fermions on
a solid-state matrix subject to random external electromagnetic
fields. In this section, we specifically consider situations wherein
the interaction between this compound system and the environment is
described by the following Hamiltonian.
\begin{eqnarray}
H\left(t\right) =
-\frac{1}{2}\mu\left[B\left(t\right)\left(\sigma^{{\rm
A}}_{z}+\sigma^{{\rm B}}_{z}\right) +b_{{\rm
A}}\left(t\right)\sigma^{{\rm A}}_{z}+b_{{\rm
B}}\left(t\right)\sigma^{{\rm B}}_{z}\right], \label{H1}
\end{eqnarray}
\noindent where, for example, $\mu$ is taken to be the gyromagnetic
ratio of the particle and $B\left(t\right)$, $b_{{\rm
A}}\left(t\right)$, and $b_{{\rm B}}\left(t\right)$ represent
stochastic environmental noise from electromagnetic fields, in
particular, with $B(t)\neq 0$; we take $\hbar= 1$ and $\sigma_{z}$
to be the well-known Pauli matrix; superscripts ${\rm A}({\rm B})$
indicate the particle Hilbert spaces $\mathcal{H}_{{\rm A}({\rm
B})}$ in which the operators act. The noise terms $B\left(t\right)$,
$b_{\rm A}\left(t\right)$, and $b_{{\rm B}}\left(t\right)$ are taken
to be statistically independent classical Markov processes
satisfying
\begin{eqnarray}
&&\left\langle B\left(t\right) \right\rangle = 0,\\
&&\left\langle B\left(t\right)B\left(t'\right) \right\rangle = \frac{\Gamma_{\rm AB}}{\mu^2}\delta\left(t-t'\right),\\
&&\left\langle b_{X}\left(t\right) \right\rangle = 0,\\
&&\left\langle b_{X}\left(t\right) b_{X}\left(t'\right)
\right\rangle = \frac{\Gamma_{X}}{\mu^2}\delta\left(t-t'\right),\ \
\ \ \ X = {\rm A},{\rm B}\ ,
\end{eqnarray}
where $\left\langle\cdots\right\rangle$ is the ensemble
time-average; $\Gamma_{X}$ denotes the phase-damping rates
associated with $b_X(t)$ $(X = {\rm A}, {\rm B})$ and $\Gamma_{AB}$
is that associated with $B(t)$. The two-qubit standard-basis
eigenstates here are
\begin{eqnarray}
&&\ket{1}_{{\rm AB}} = \ket{++}_{{\rm AB}}, \ket{2}_{{\rm AB}} = \ket{+-}_{{\rm AB}},\nonumber\\
&&\ket{3}_{{\rm AB}} = \ket{-+}_{{\rm AB}}, \ket{4}_{{\rm AB}} = \ket{--}_{{\rm AB}}.
\label{2QubitBasis}
\end{eqnarray}
The Hamiltonian of Eq. \ref{H1}, the same one assumed in
\cite{YE03}, is representative of the class of interactions giving
rise to pure dephasing decoherence. The corresponding dynamical
process is described here as a ``quantum channel,'' in accord with
the standard terminology of qubit evolution as described by linear
maps \cite{NC00,JQ06}, namely, a phase-damping channel, using the
operator-sum representation which we now introduce.

The time-dependent density matrix for the two-qubit system can be
obtained by taking ensemble averages over the three noise fields,
$B\left(t\right)$, $b_{A}\left(t\right)$, and $b_{{\rm
B}}\left(t\right)$, that is,
\begin{eqnarray}
\rho\left(t\right) = \left\langle\left\langle\left\langle \rho_{\rm
st}\left(t\right) \right\rangle\right\rangle\right\rangle,
\label{rhoAverage}
\end{eqnarray}
where the statistical density operator $\rho_{\rm st}\left(t\right)$
and the unitary operator $U\left(t\right)$ associated with $H(t)$
are
\begin{eqnarray}
\rho_{\rm st}\left(t\right) =
U\left(t\right)\rho\left(0\right)
U^{\dagger}\left(t\right)\ \ \ {\rm and}\\
U\left(t\right) = \exp\left[-i\int_{0}^{t}{dt'
H\left(t'\right)}\right]\ ,
\end{eqnarray}
respectively. It is helpful to consider the dynamical evolution of
$\rho(t)$ as a completely positive trace preserving (CPTP) linear
map $\mathcal{E}(\rho)$, that is, a combination of local and
collective quantum channels, any of which can be turned off in
particular cases, taking an input state $\rho\left(0\right)$ to the
output state $\rho\left(t\right)$ given by the operator sum
\begin{eqnarray}
\rho\left(t\right) = \mathcal{E}\left(\rho\left(0\right)\right) =
\sum_{\mu =
1}^{N}\overline{E}_{\mu}^{\dagger}\left(t\right)\rho\left(0\right)
\overline{E}_{\mu}\left(t\right)\label{kraussSumDefinition},
\end{eqnarray}
where $\overline{E}_{\mu}$ are decomposition operators that satisfy
the completeness relation
\begin{eqnarray}
\sum_{\mu}\overline{E}_{\mu}^{\dagger}\overline{E}_{\mu} =
\mathbb{I}\ . \label{kraussCompleteness}
\end{eqnarray}
In each of the various cases considered here, the internal structure
of the $\overline{E}_{\mu}$ will be accord with the Hamiltonian;
various terms may or may not nontrivially contribute in a given
case. The most general solution of Eq. \ref{kraussSumDefinition},
assuming that the system is not initially correlated with any of the
three environments, is
\begin{eqnarray}
\rho\left(t\right) = \sum_{i,j=1}^{2}\sum_{k=1}^{3}
\big(D_{k}^{\dagger}E_{j}^{{\rm B}\dagger}E_{i}^{{\rm
A}\dagger}\big)\rho\left(0\right)\big(E_{i}^{{\rm A}}E_{j}^{{\rm
B}}D_{k}\big),
\end{eqnarray}
where the terms describing the interaction with the local magnetic
fields $b_{{\rm A}}\left(t\right)$ and $b_{{\rm B}}\left(t\right)$
involve the decomposition operators
\begin{eqnarray}
E_{1}^{{\rm A}} = \left(
\begin{array}{ccc}
 1 & 0\\
 0 & \gamma_{{\rm A}}\left(t\right)
\end{array}
\right) \otimes\mathbb{I},\ \  E_{2}^{{\rm A}} = \left(
\begin{array}{ccc}
 0 & 0\\
 0 & \omega_{{\rm A}}\left(t\right)
\end{array}
\right) \otimes\mathbb{I},
\\
\nonumber\\
E_{1}^{{\rm B}} =\mathbb{I} \otimes \left(
\begin{array}{ccc}
 1 & 0\\
 0 & \gamma_{{\rm B}}\left(t\right)
\end{array}
\right),\ \ E_{2}^{{\rm B}} =\mathbb{I} \otimes \left(
\begin{array}{ccc}
 0 & 0\\
 0 & \omega_{{\rm B}}\left(t\right)
\end{array}
\right).
\end{eqnarray}

\noindent By contrast, the operators describing the collective
interaction with a non-zero global magnetic field,
$B\left(t\right)$, are
\begin{eqnarray}
D_{1} = \left(
\begin{array}{cccc}
\gamma\left(t\right) & 0 & 0 & 0 \\
0 & 1 & 0 & 0 \\
0 & 0 & 1 & 0 \\
0 & 0 & 0 & \gamma\left(t\right)
\end{array}\label{D_{1}}
\right),
\\
D_{2} = \left(
\begin{array}{cccc}
 \omega_{1}\left(t\right) & 0 & 0 & 0 \\
 0 & 0 & 0 & 0 \\
 0 & 0 & 0 & 0 \\
 0 & 0 & 0 & \omega_{2}\left(t\right)
\end{array}
\right),\label{D_{2}}
\\
D_{3} = \left(
\begin{array}{cccc}
 0 & 0 & 0 & 0 \\
 0 & 0 & 0 & 0 \\
 0 & 0 & 0 & 0 \\
 0 & 0 & 0 & \omega_{3}\left(t\right)
\end{array}\label{D_{3}}
\right),
\end{eqnarray}
\noindent \cite{YE03}. The time-dependent parameters appearing in
Eqs. 13--17 above are
\begin{eqnarray}
\gamma_{{\rm A}}\left(t\right) = e^{-t/2T_{{\rm A}}},
\gamma_{{\rm B}}\left(t\right) = e^{-t/2T_{{\rm B}}},\\
\omega_{{\rm A}}\left(t\right) = \sqrt{1-\gamma_{{\rm A}}^{2}},
\omega_{{\rm B}}\left(t\right) = \sqrt{1-\gamma_{{\rm B}}^{2}},\\
\gamma\left(t\right) = e^{-t/2T_{{\rm AB}}},\\
\omega_{1}\left(t\right) = \sqrt{1-\gamma^{2}},\\
\omega_{2}\left(t\right) = -\gamma^{2}\sqrt{1-\gamma^{2}},\\
\omega_{3}\left(t\right) = \sqrt{\left(1-\gamma^{2}\right)
\left(1-\gamma^{4}\right)},
\end{eqnarray}
where $T_X=\frac{1}{\Gamma_X}\ \ (X = {\rm A}, {\rm B}, {\rm AB}) $
are the phase-relaxation times introduced in \cite{YE03}, $\Gamma_X$
being the rate parameters appearing on the right-hand sides of Eqs.
3 and 5; from here on, for tractability of notation, time does not
explicitly appear as an argument for these quantities but is
implied.

The (two-qubit) entanglement is measured here by the concurrence
\begin{eqnarray}
C\left(\rho_{AB}\right) = \max \left[0, \sqrt{\lambda_{1}} -
\sqrt{\lambda_{2}} - \sqrt{\lambda_{3}} - \sqrt{\lambda_{4}}\right]\
,\label{concurrenceDefinition}
\end{eqnarray}
where $\lambda_{i}$ are the eigenvalues of the matrix
\begin{eqnarray}
\rho_{AB}\tilde{\rho}_{AB} \equiv \rho_{AB}\left(\sigma_{y}^{{\rm
A}}\otimes\sigma_{y}^{{\rm B}}\right) \rho_{AB}^{\ast}
\left(\sigma_{y}^{{\rm A}}\otimes\sigma_{y}^{{\rm B}} \right)
\end{eqnarray}
indexed according to decreasing magnitude, $\rho_{AB}^{\ast}$
denotes the complex conjugate of $\rho_{AB}$, and $\sigma_{y}^{{\rm
A}({\rm B})}$ is the standard Pauli matrix acting in the space of
qubit A(or B) \cite{W98}. Concurrence is related to the canonical
measure of entanglement, the entanglement of formation, ``$E_{f}$'',
by
\begin{eqnarray}
E_{f}\left(\rho\right) =
h\left(\frac{1 + \sqrt{1-C^{2}\left(\rho\right)}}{2}\right),
\end{eqnarray}
where $h\left(x\right) = -x\log_{2}{x} -
\left(1-x\right)\log_{2}{\left(1-x\right)}$, of which it is,
therefore, a monotonic function.

\subsection{\label{two-qubitResults}DECOHERENCE AND DISENTANGLEMENT}

In the standard-basis representation of Eq. 6, the generic pure
state of the two-qubit system is
\begin{eqnarray}
\ket{\Psi} = a\ket{1}_{{\rm AB}} + b\ket{2}_{{\rm AB}} +
c\ket{3}_{{\rm AB}} + d\ket{4}_{{\rm AB}},
\end{eqnarray}
a normalized state-vector with $a,b,c,d\in \mathbb{C}$. The generic
class of two-qubit pure states represented by $|\Psi\rangle$
contains two pertinent subclasses, distinguished by the large-time
behavior of their coherence under collective dephasing noise, that
is, dephasing in which each qubit interacts with the same global
noise field $B(t)$: one class is fragile, whereas the other is
robust \cite{YE03}.

\vskip 3pt

\noindent (i) The \textit{fragile} class has the forms
\begin{eqnarray}
\ket{\phi_{1}} = a\ket{1}_{{\rm AB}} + b\ket{2}_{{\rm AB}} + d\ket{4}_{{\rm AB}}\ ,\\
\ket{\phi_{2}} = a\ket{1}_{{\rm AB}} + c\ket{3}_{{\rm AB}} + d\ket{4}_{{\rm AB}}\ .
\end{eqnarray}
\noindent (ii) The \textit{robust} class has the forms
\begin{eqnarray}
\ket{\psi_{1}} = a\ket{1}_{{\rm AB}} + b\ket{2}_{{\rm AB}} + c\ket{3}_{{\rm AB}}\ ,\\
\ket{\psi_{2}} = b\ket{2}_{{\rm AB}} + c\ket{3}_{{\rm AB}} + d\ket{4}_{{\rm AB}}\ .
\end{eqnarray}
Without loss of generality, we consider the first of the two forms
in each case, for the purposes of precisely characterizing the
decoherence and disentanglement behavior of these classes.

The fragile states have been shown to decohere under collective
dephasing noise in such a way that \emph{all} off-diagonal matrix
elements go to zero exponentially as a function of time. A well
known example of the fragile class is that with $a=d=1/\sqrt{2},
b=c=0$, namely, the spin-triplet Bell state $|\Phi^+\rangle$. By
contrast, the robust states remain partially coherent under
collective noise, some off-diagonal elements being constants
\cite{LW03}. A well known example of the robust class is that with
$a=d=0, b=-c=1/\sqrt{2}$, the spin-singlet Bell state
$|\Psi^-\rangle$, which is known to be decoherence-free under
collective dephasing \cite{Kwiat}. In fact, each class of states
contains two Bell states.

In \cite{YE03}, the evolution of state entanglement under local
dephasing noise for both the fragile and robust classes was
characterized; it was shown for both classes, for both the case of
single-qubit local and the case of multi-local dephasing noise, that
the disentanglement rate can be much faster than the decoherence
rate when either (and so both) occur. We now generalize and sharpen
this result, by pointing out that the decoherence timescale bounds
the disentanglement timescale from above and explicitly showing that
such a bounding relation also holds under collective dephasing
noise, {\it i.e.}, when $B\left(t\right)\neq 0$, a case in which
disentanglement was not explicitly investigated in \cite{YE03} and
only to a limited extent and in a somewhat different manner and
context in the preceding study \cite{YE02}. Here, we consider all
possible cases in an identical way in a common formalism (that in
subsequent sections is extended to the three-qubit context). In this
way, we confirm and sharpen the general conclusions for two-qubit
systems drawn in \cite{YE03} before proceeding on to consider the
situation of central interest here, namely, three-qubits subject to
dephasing noise.

The initial-time density matrix for the representative fragile-state
class is
\begin{eqnarray}
\rho^{\rm F}_{{\rm AB}}\left(t=0\right)=P(\ket{\phi_1})\equiv \left(
\begin{array}{cccc}
 |a|^2 & a\:b^{\ast} & 0 & a\:d^{\ast} \\
 b\:a^{\ast} & |b|^2 & 0 & b\:d^{\ast} \\
 0 & 0 & 0 & 0 \\
 d\:a^{\ast} & d\:b^{\ast} & 0 & |d|^2
\end{array}\label{fragileInitial}
\right),
\end{eqnarray}
whereas for the robust class one has
\begin{eqnarray}
\rho^{\rm R}_{{\rm AB}}\left(t=0\right) =P(\ket{\psi_1})\equiv\left(
\begin{array}{cccc}
 |a|^2 & a\:b^{\ast} & a\:c^{\ast} & 0\\
 b\:a^{\ast} & |b|^2 & b\:c^{\ast} & 0\\
 c\:a^{\ast} & c\:b^{\ast} & |c|^2 & 0\\
 0 & 0 & 0 & 0
\end{array}\label{robustInitial}
\right)\ ,
\end{eqnarray}
$P(\ket{\upsilon})$ indicating the projector corresponding to the
generic state-vector argument $\ket{\upsilon}$ \cite{YE03}.

The time-evolved density matrix of a system beginning at $t=0$ in
the fragile-class state under the collective dephasing channel
described by Eqs. \ref{D_{1}}--\ref{D_{3}} is given by
$\rho\left(t\right) = \mathcal{E}\left(\rho\left(0\right)\right) =
\sum_{k = 1}^{3}D_{k}^{\dagger}\rho\left(0\right)D_{k}$:
\begin{eqnarray}
\rho^{\rm F}_{{\rm AB}}\left(t\right) = \left(
\begin{array}{cccc}
 |a|^{2} & a\:b^{\ast}\gamma  & 0 & a\:d^{\ast}\gamma^{4} \\
 b\:a^{\ast}\gamma  & |b|^{2} & 0 & b\:d^{\ast}\gamma  \\
 0 & 0 & 0 & 0 \\
 d\:a^{\ast}\gamma^{4} & d\:b^{\ast}\gamma & 0 & \left|d\right|^{2}
\end{array}
\right)\ ,
\end{eqnarray}
which exhibits two pertinent two-qubit decoherence timescales,
parameterized by $\gamma^{4}$ (fast) and $\gamma$ (slow), and
becomes diagonal in the limit of large times. By comparison with the
result in the independent multi-local noise case (see Eq. 64 of
\cite{YE03}) wherein two timescales of decoherence exist that are
determined by $\gamma_{{\rm A}({\rm B})}$ and $\gamma_{{\rm
A}}\gamma_{{\rm B}}$, the decay of some off-diagonal terms in this
case is more rapid, being determined by $\gamma^4$.

The robust states $\rho^{\rm R}_{{\rm AB}}(0)$ under collective
dephasing noise become
\begin{eqnarray}
\rho^{\rm R}_{{\rm AB}}\left(t\right) = \left(
\begin{array}{cccc}
 |a|^{2} & a\:b^{\ast}\gamma  & a\:c^{\ast}\gamma  & 0 \\
 b\:a^{\ast}\gamma  & |b|^{2} & b\:c^{\ast} & 0 \\
 c\:a^{\ast}\gamma  & c\:b^{\ast} & |c|^{2} & 0 \\
 0 & 0 & 0 & 0
\end{array}
\right)\ ,
\end{eqnarray}
and, by contrast, do \emph{not} become diagonal in the large-time
limit. Furthermore, also in contrast to the case of the fragile
class, we see that the off-diagonal elements that do decay do so
according to only a single, slow timescale, namely, that given by a
single factor of $\gamma$. The robustness of this class is due to
the fact that the off-diagonal elements $\rho_{23}$ and $\rho_{32}$
(corresponding to the symmetric spin-singlet component \cite{Kwiat})
do not decay; the robust states remain partially coherent in the
limit of large time. By contrast, under multi-local dephasing (see
Eq. 67 of \cite{YE03}), density matrix elements $\rho_{23}$ and
$\rho_{32}$ \emph{do} decay to zero, in the timescale determined by
$\gamma$. Under collective dephasing noise, the two-qubit system
coherence-decay timescales determined by the $\gamma$ factors in the
above density matrices, are thus
\begin{eqnarray}
\tau_{\rm 2-dec}^{\rm F\left({\rm slow}\right)}&=& 2\bigg({1\over\Gamma_{\rm AB}}\bigg)\ ,\\
\tau_{\rm 2-dec}^{\rm F\left({\rm fast}\right)}&=& {1\over 2}\bigg({1\over\Gamma_{\rm AB}}\bigg)\ ,\\
\tau_{\rm 2-dec}^{\rm R}&=&2\bigg({1\over\Gamma_{\rm AB}}\bigg)\ \ \
({\rm for\ decaying\ elements)}.
\end{eqnarray}

Now consider \emph{single-qubit} decoherence under collective
dephasing noise, corresponding to the reduction of magnitude of
off-diagonal elements of the single-qubit reduced density matrices.
Because the action of $B\left(t\right)$ on each qubit is identical,
it suffices to consider the reduced density matrix of just one qubit
(here, we choose A) for both the fragile and robust classes. One
finds
\begin{eqnarray}
\rho^{\rm F}_{{\rm A}} = {\rm tr}_{{\rm B}}\rho_{{\rm AB}}^{\rm
F}\left(t\right) = \left(
\begin{array}{ccc}
 |a|^{2}+\left|b\right|^{2} & b\:d^{\ast}\gamma\\
 d\:b^{\ast}\gamma & |d|^{2}
\end{array}
\right),
\\
\rho^{\rm R}_{{\rm A}} = {\rm tr}_{{\rm A}}\rho_{{\rm AB}}^{\rm
R}\left(t\right) = \left(
\begin{array}{ccc}
 |a|^{2}+|b|^{2} & a\:c^{\ast}\gamma \\
 c\:a^{\ast}\gamma & |c|^{2}
\end{array}
\right).
\end{eqnarray}
\noindent For the two classes, single-qubit coherence is entirely
lost as parameterized by $\gamma$, namely, in the slow decoherence
time-scale given above, so the single-qubit decoherence timescales
are the same:
\begin{eqnarray}
\tau_{\rm 1-dec}^{\rm F}&=& \bigg(\frac{1}{\Gamma}\bigg)\ ,\\
\tau_{\rm 1-dec}^{\rm R}&=& \bigg(\frac{1}{\Gamma}\bigg)\ .
\end{eqnarray}

Quantum state disentanglement is described by a reduction of
concurrence, $C(\rho)$. For fragile states, one finds
\begin{eqnarray}
C\left(\rho_{{\rm AB}}^{\rm F}\left(t\right)\right) = 2
\gamma^{4}\left|a\right| \left|d\right|.
\end{eqnarray}
Disentanglement of these states is therefore characterized by the
time
\begin{eqnarray}
\tau_{\rm dis}={1\over 2}\bigg({1\over{\Gamma_{AB}}}\bigg)\ ,
\end{eqnarray}
that is, the \emph{fast} time-scale for two-qubit decoherence above,
{\it cf.} Eq. 37. A factor of $\gamma^{4}$ appears here, by contrast
to the case of multi-local dephasing (see Eq. 66 of \cite{YE03})
where disentanglement takes place via the factor $\gamma_{{\rm
A}}\gamma_{{\rm B}}$. Under collective dephasing, disentanglement
thus occurs more quickly than in multi-local dephasing due to an
additional factor of $\gamma^2$ in off-diagonal terms (assuming
$\gamma_{AB},\gamma_{A}$ and $\gamma_{B}$ to be comparable in value,
which enables a meaningful comparison of effects). For the robust
class of states, the entanglement is
\begin{eqnarray}
C\left(\rho_{{\rm AB}}^{\rm R}\left(t\right)\right) = 2
\left|b\right| \left|c\right|\ ;
\end{eqnarray}
that is, there is \emph{no} disentanglement for the robust class,
which also does not decohere. We therefore see that this class of
state can be identified on the basis of the robustness of its
entanglement, as well as that of its coherence. In the following
section, we find a similar behavior distinguishing the two major
entanglement classes of three-qubit states.

Thus, the relation between the timescales of decoherence and
disentanglement described in \cite{YE03} continues to hold under the
collective dephasing channel, in which $B\left(t\right) \neq 0$ and
$b_X(t)=0$. With the above results, together with those of
\cite{YE03}, one now can consider the behavior of entanglement under
dephasing noise to be fully characterized for both these
complementary classes of two-qubit system states, because both
local-noise and collective-noise cases have been fully analyzed. In
the following section, we consider three-qubit composite systems
under local and collective dephasing noise, which involves a greater
number of scenarios involving combinations of these noise types.


\section{\label{three-QubitSystem}THREE-QUBIT SYSTEM}
Now consider the effect of similar dephasing noise on states of
three qubits. In three-qubit systems (ABC) there are known to be two
qualitatively different pure-state entanglement classes,
\begin{eqnarray}
\ket{{\rm W^g}}&=&\bar{a}_{1}\ket{001} + \bar{a}_{2}\ket{010} + \bar{a}_{4}\ket{100}\ ,\nonumber\\
\ket{{\rm GHZ^g}}&=&\bar{a}_{0}\ket{000} + \bar{a}_{7}\ket{111},
\label{WandGHZ}
\end{eqnarray}
\noindent with coefficients $\bar{a}_{i}\in\mathbb{C}$ such that
$\left|\bar{a}_{1}\right|^{2} + \left|\bar{a}_{2}\right|^{2} +
\left|\bar{a}_{4}\right|^{2} = 1$, and $\left|\bar{a}_{0}\right|^{2}
+  \left|\bar{a}_{7}\right|^{2} = 1$, defined by interconvertibility
under stochastic local operations and classical communication
(SLOCC) operations \cite{DVC00}. We will accordingly consider each
class individually under the available combinations of local and
collective dephasing-noise that this system might encounter,
assuming that a single noise type affects each qubit. In the above
classification of (here, initial) states, the distinguishing
property of the W class is, equivalently, the robustness of its
internal entanglement under the loss of any one qubit of the system,
in the sense of retaining the entanglement of the remaining two
qubits. By contrast, the GHZ class has maximum three-qubit
entanglement \cite{CKW00} but no two-qubit entanglement after the
loss of any qubit.


\subsection{\label{three-QubitModel}MODEL AND MEASURES}

We now consider the three-qubit system ABC in dephasing-noise
environments that may act at the level of each of the three
individual single-qubit subsystems ($X=$ A, B, C) and/or the three
two-qubit subsystems ($XY=$ AB, AC, and BC), or the full three-qubit
system (ABC) alone. Thus, the noise terms considered here are those
due to the local-dephasing environment which acts on a single qubit,
$B^{(1)}$, a two-qubit collective-dephasing environment, $B^{(2)}$,
and a three-qubit collective-dephasing environment, $B^{(3)}$.  We
consider only cases in which a given qubit is influenced by a single
type of noise, in order to clearly distinguish the influence of
combinations of dephasing noise on three-qubit systems.

The corresponding three-qubit interaction Hamiltonian is
\begin{eqnarray}
H\left(t\right)&=&-\frac{1}{2}\mu\bigg[B^{(1)}_{{\rm
A}}\left(t\right)\sigma^{{\rm A}}_{z}+ B^{(1)}_{{\rm
B}}\left(t\right)\sigma^{{\rm B}}_{z}+
B^{(1)}_{{\rm C}}\left(t\right)\sigma^{{\rm C}}_{z}\nonumber \\
&&+\:B^{(2)}_{{\rm AB}}\left(t\right) \left(\sigma^{{\rm A}}_{z} + \sigma^{{\rm B}}_{z}\right) \nonumber \\
&&+\:B^{(2)}_{{\rm BC}}\left(t\right) \left(\sigma^{{\rm B}}_{z} + \sigma^{{\rm C}}_{z}\right) \nonumber \\
&&+\:B^{(2)}_{{\rm AC}}\left(t\right) \left(\sigma^{{\rm A}}_{z} + \sigma^{{\rm C}}_{z}\right) \nonumber \\
&&+\:B^{(3)}\left(t\right) \left(\sigma^{{\rm A}}_{z} + \sigma^{{\rm
B}}_{z} + \sigma^{{\rm C}}_{z}\right)\bigg]\ , \label{H3}
\end{eqnarray}
where $\hbar$ is taken to be unity, and for example, $\mu$ is the
gyromagnetic ratio and the $B^{(i)}\ (i=1,2,3)$ are the imposed
stochastic magnetic fields at each of the three available scales,
with subscripts indicating the region of influence in cases in which
only subsystems are affected; Pauli matrices are also labeled by the
subsystem on which they act. The standard three-qubit Hilbert space
basis assumed here is
\begin{eqnarray}
&\ket{1}_{{\rm ABC}} = \ket{000}_{{\rm ABC}}, \ket{2}_{{\rm ABC}} = \ket{001}_{{\rm ABC}}\ ,\nonumber\\
&\ket{3}_{{\rm ABC}} = \ket{010}_{{\rm ABC}}, \ket{4}_{{\rm ABC}} = \ket{011}_{{\rm ABC}}\ ,\nonumber\\
&\ket{5}_{{\rm ABC}} = \ket{100}_{{\rm ABC}}, \ket{6}_{{\rm ABC}} = \ket{101}_{{\rm ABC}}\ ,\nonumber\\
&\ket{7}_{{\rm ABC}} = \ket{110}_{{\rm ABC}}, \ket{8}_{{\rm ABC}} =
\ket{111}_{{\rm ABC}}\ ,
\end{eqnarray}
where $\ket{ijk}_{{\rm ABC}} \equiv \ket{i}_{{\rm A}} \otimes
\ket{j}_{{\rm B}} \otimes
 \ket{k}_{{\rm C}}$ ($i,j,k=0,1$) denote the eigenstates of the product operator
$\sigma^{{\rm A}}_{z} \otimes \sigma^{{\rm B}}_{z} \otimes
\sigma^{{\rm C}}_{z}$. For simplicity, much as before, we assume
that the stochastic fields are taken to be classical and
characterized as statistically independent Markov processes
satisfying

\begin{eqnarray}
\hspace{-5.0 cm}
&&\left\langle B^{(1)}_{X}\left(t\right) \right\rangle = 0,\\
\nonumber\\
&&\left\langle B^{(1)}_{X}\left(t\right)B^{(1)}_{X}\left(t'\right)
\right\rangle =
 \frac{\Gamma_{{\rm 1}}}{\mu^{2}}\delta\left(t-t'\right),\\
\nonumber\\
&&\left\langle B^{(2)}_{XY}\left(t\right) \right\rangle = 0,\\
\nonumber\\
&&\left\langle B^{(2)}_{XY}\left(t\right)B^{(2)}_{XY}\left(t'\right)
\right\rangle
= \frac{\Gamma_{{\rm 2}}}{\mu^{2}}\delta\left(t-t'\right),\\
\nonumber\\
&&\left\langle B^{(3)}\left(t\right) \right\rangle = 0,\\
\nonumber\\
&&\left\langle B^{(3)}\left(t\right)B^{(3)}\left(t'\right)
\right\rangle = \frac{\Gamma_{{\rm
3}}}{\mu^2}\delta\left(t-t'\right) \ ,
\end{eqnarray}
where $\left\langle \cdots \right\rangle$ stands for ensemble
average. Here $\Gamma_i$ is the damping rate due, for example, to
the respective interaction with magnetic fields localized to
$i=1,2,3$ qubits at a time. The imposed white-noise conditions on
the three-qubit system and any subsystems ensure a Markovian
time-evolution. They also require interactions at a given scale to
be of the same strength, which allows one to make an objective
assessment of the importance of interactions by scale, as is done
below.

The compound-system density matrix can be obtained by taking the
ensemble averages over the three noise sources, which are again
given by decomposition-operators. For the single-qubit dephasing
channel, in this section labeled $\mathcal{D}$, we have the
following decomposition-operator matrices for qubit A at the
one-qubit level.

\begin{eqnarray}
D^{\rm A}_{1} =
\left(
\begin{array}{ccc}
 1 & 0\\
 0 & \gamma_{{\rm A}}\left(t\right)
\end{array}
\right)
\otimes
\left(
\begin{array}{ccc}
 1 & 0\\
 0 & 1
\end{array}
\right) \otimes \left(
\begin{array}{ccc}
 1 & 0\\
 0 & 1
\end{array}
\right).
\label{A_{1}}
\\
D^{\rm A}_{2} =
\left(
\begin{array}{ccc}
 0 & 0\\
 0 & \omega_{{\rm A}}\left(t\right)
\end{array}
\right)
\otimes
\left(
\begin{array}{ccc}
 1 & 0\\
 0 & 1
\end{array}
\right)
\otimes
\left(
\begin{array}{ccc}
 1 & 0\\
 0 & 1
\end{array}
\right).
\label{A_{2}}
\end{eqnarray}
(For the dephasing of qubits B and C, the matrices are analogous.
Note also that the choice of decomposition operator labels in this
section differs from that of the previous section, which there was
made to agree with the notation of \cite{YE03}.) As before, the time
parameter in $\gamma$s and $\omega$s will be implicit from here on.

For the collective dephasing channel that acts only on two-qubit
subsystems, in this section labeled $\mathcal{E}$, we have the
following decomposition-operator matrices for the AB qubit system at
the two-qubit level, which are just those that are collective at the
two-qubit level. (For collective-dephasing-noise contributions
acting the two-qubit systems BC and AC, the decomposition operator
matrices are analogous.)
\begin{eqnarray}
E^{\rm AB}_{1} = \left(
\begin{array}{cccc}
\gamma_{\rm AB} & 0 & 0 & 0 \\
0 & 1 & 0 & 0 \\
0 & 0 & 1 & 0 \\
0 & 0 & 0 & \gamma_{\rm AB}
\end{array}
\right)
\otimes
\left(
\begin{array}{ccc}
 1 & 0\\
 0 & 1
\end{array}
\right).
\\
E^{\rm AB}_{2} = \left(
\begin{array}{cccc}
 \omega_{\rm AB1} & 0 & 0 & 0 \\
 0 & 0 & 0 & 0 \\
 0 & 0 & 0 & 0 \\
 0 & 0 & 0 & \omega_{\rm AB2}
\end{array}
\right)
\otimes
\left(
\begin{array}{ccc}
 1 & 0\\
 0 & 1
\end{array}
\right).
\\
E^{\rm AB}_{3} = \left(
\begin{array}{cccc}
 0 & 0 & 0 & 0 \\
 0 & 0 & 0 & 0 \\
 0 & 0 & 0 & 0 \\
 0 & 0 & 0 & \omega_{\rm AB3}
\end{array}
\right)
\otimes
\left(
\begin{array}{ccc}
 1 & 0\\
 0 & 1
\end{array}
\right).
\end{eqnarray}
Finally, for the three-qubit collective dephasing channel, we have
the following decomposition-operator matrices acting on the whole
three-qubit system.
\begin{eqnarray}
F^{\rm ABC}_{1} =
\left(
\begin{array}{cccccccc}
 \gamma_{\rm ABC} & 0 & 0 & 0 & 0 & 0 & 0 & 0\\
 0 & 1 & 0 & 0 & 0 & 0 & 0 & 0\\
 0 & 0 & 1 & 0 & 0 & 0 & 0 & 0\\
 0 & 0 & 0 & 1 & 0 & 0 & 0 & 0\\
 0 & 0 & 0 & 0 & 1 & 0 & 0 & 0\\
 0 & 0 & 0 & 0 & 0 & 1 & 0 & 0\\
 0 & 0 & 0 & 0 & 0 & 0 & 1 & 0\\
 0 & 0 & 0 & 0 & 0 & 0 & 0 & \gamma_{{\rm ABC}}
\end{array}
\right).
\\
F^{\rm ABC}_{2} =
\left(
\begin{array}{cccccccc}
 \omega_{\rm ABC1} & 0 & 0 & 0 & 0 & 0 & 0 & 0\\
 0 & 0 & 0 & 0 & 0 & 0 & 0 & 0\\
 0 & 0 & 0 & 0 & 0 & 0 & 0 & 0\\
 0 & 0 & 0 & 0 & 0 & 0 & 0 & 0\\
 0 & 0 & 0 & 0 & 0 & 0 & 0 & 0\\
 0 & 0 & 0 & 0 & 0 & 0 & 0 & 0\\
 0 & 0 & 0 & 0 & 0 & 0 & 0 & 0\\
 0 & 0 & 0 & 0 & 0 & 0 & 0 & \omega_{{\rm ABC2}}
\end{array}
\right).
\\
F^{\rm ABC}_{3} =
\left(
\begin{array}{cccccccc}
 0 & 0 & 0 & 0 & 0 & 0 & 0 & 0\\
 0 & 0 & 0 & 0 & 0 & 0 & 0 & 0\\
 0 & 0 & 0 & 0 & 0 & 0 & 0 & 0\\
 0 & 0 & 0 & 0 & 0 & 0 & 0 & 0\\
 0 & 0 & 0 & 0 & 0 & 0 & 0 & 0\\
 0 & 0 & 0 & 0 & 0 & 0 & 0 & 0\\
 0 & 0 & 0 & 0 & 0 & 0 & 0 & 0\\
 0 & 0 & 0 & 0 & 0 & 0 & 0 & \omega_{{\rm ABC3}}
\end{array}
\right).
\end{eqnarray}

The parameters appearing in the previous matrices are:
\begin{eqnarray}
\gamma_{i} = e^{-t/2T_{i}},\\
\omega_{i} = \sqrt{1 - e^{-t/T_{i}}} = \sqrt{1-\gamma_{i}^{2}},\\
\omega_{i1} = \sqrt{1-\gamma_{i}^{2}},\\
\omega_{i2} = -\gamma_{i}^{2}\sqrt{1-\gamma_{i}^{2}},\\
\omega_{i3} = \sqrt{\left(1-\gamma_{i}^{2}\right)
\left(1-\gamma_{i}^{4}\right)}.
\end{eqnarray}
Here, as before, $T_i = \frac{1}{\Gamma_i}$ is the phase-relaxation
time associated with the corresponding interaction. Below, the
numeric labels $i$ may be replaced by the labels of the pertinent
qubits A, AB, ABC, and so on, for specificity.

The explicit form of the time-evolved density matrix subject to the
available one-qubit, two-qubit, and three-qubit dephasing noise
sources is given by
\begin{widetext}
\begin{eqnarray}
\rho\left(t\right) = \sum_{i,j,k=1}^{2}\sum_{\ l,m,n,p=1}^{3}
\Big(F_{p}^{\rm ABC\dagger} E_{n}^{{\rm AC}\dagger} E_{m}^{\rm
BC\dagger}E_{l}^{\rm AB\dagger}D_{k}^{\rm C\dagger}D_{j}^{\rm
B\dagger}D_{i}^{\rm A\dagger}\Big) \rho\left(0\right)\Big(D^{\rm
A}_{i}D^{\rm B}_{j}D^{\rm C}_{k}E_{l}^{\rm AB} E_{m}^{\rm BC}
E_{n}^{\rm AC}F^{\rm ABC}_{p}\Big)\label{timeEvolver}.
\end{eqnarray}
\end{widetext}
For each sort of dephasing-noise environment considered, the
relevant off-diagonal elements from the single-qubit-reduced,
two-qubit-reduced, and three-qubit density matrices exhibit any
one-qubit, two-qubit, and three-qubit decoherence effects,
respectively, and bipartite disentanglement occurs when there is a
reduction of $C_{{\rm (XY)}}^{2}$, defined in Eq.
\ref{concurrenceDefinition}, where $X,Y={\rm A,B,C}$. (See the note
\cite{Note1} for a discussion of the grounds for this choice of
entanglement measure.)

\subsection{\label{behaviorW}BEHAVIOR OF W-CLASS STATES}

Here, the environmental noise channel inducing single-qubit
local-dephasing on qubit A is given explicitly; the single-qubit
local dephasing noise acting on qubit B and qubit C, respectively,
are of the same general form. In the case of subsystem-collective
dephasing noise, considered in Subsection 2 below, where the
collective-noise channel acting on qubits A and B is discussed, a
similar statement applies for the channels acting on B and C only
and A and C only. The initial generalized W-class state density
matrix
\begin{eqnarray}
\rho^{{\rm W}}_{{\rm ABC}}\left(0\right)=P(\ket{\rm W^g}) \equiv
\left(
\begin{array}{cccccccc}
 0 & 0 & 0 & 0 & 0 & 0 & 0 & 0 \\
 0 & \left|\bar{a}_{1}\right|^{2} & \bar{a}_{1}\bar{a}_{2}^{\ast} & 0 & \bar{a}_{1}\bar{a}_{4}^{\ast} & 0 & 0 & 0 \\
 0 & \bar{a}_{2}\bar{a}_{1}^{\ast} & \left|\bar{a}_{2}\right|^{2} & 0 & \bar{a}_{2}\bar{a}_{4}^{\ast} & 0 & 0 & 0 \\
 0 & 0 & 0 & 0 & 0 & 0 & 0 & 0 \\
 0 & \bar{a}_{4}\bar{a}_{1}^{\ast} & \bar{a}_{4}\bar{a}_{2}^{\ast} & 0 & \left|\bar{a}_{4}\right|^{2} & 0 & 0 & 0 \\
 0 & 0 & 0 & 0 & 0 & 0 & 0 & 0 \\
 0 & 0 & 0 & 0 & 0 & 0 & 0 & 0 \\
 0 & 0 & 0 & 0 & 0 & 0 & 0 & 0
\end{array}
\right).\
\end{eqnarray}
We now examine the behavior of the time-evolved state $\rho(t)$,
given by Eq. \ref{timeEvolver} above, in detail.

\subsubsection{\label{W1}ONE-QUBIT LOCAL DEPHASING CHANNEL: $\mathcal{D}$}

Under local dephasing noise,
\begin{eqnarray}
\rho_{{\rm ABC}}^{{\rm W}}\left(t\right)=
\left(
\begin{small}
\begin{array}{cccccccc}
 0 & 0 & 0 & 0 & 0 & 0 & 0 & 0 \\
 0 & |\bar{a}_{1}|^2 & \bar{a}_{1}\bar{a}_{2}^* & 0 & \bar{a}_{1}\bar{a}_{4}^* \gamma_{\rm A} & 0 & 0 & 0 \\
 0 & \bar{a}_{2}\bar{a}_{1}^* & |\bar{a}_{2}|^2 & 0 & \bar{a}_{2}\bar{a}_{4}^* \gamma_{\rm A} & 0 & 0 & 0 \\
 0 & 0 & 0 & 0 & 0 & 0 & 0 & 0 \\
 0 & \bar{a}_{4}\bar{a}_{1}^* \gamma_{\rm A} & \bar{a}_{4}\bar{a}_{2}^* \gamma_{\rm A} & 0 & |\bar{a}_{4}|^2 & 0 & 0 & 0 \\
 0 & 0 & 0 & 0 & 0 & 0 & 0 & 0 \\
 0 & 0 & 0 & 0 & 0 & 0 & 0 & 0 \\
 0 & 0 & 0 & 0 & 0 & 0 & 0 & 0
\end{array}\label{WA3Qubit}
\end{small}
\right)
\end{eqnarray}
is the time-evolved three-qubit density matrix obtained when
beginning at $t=0$ with a pure W-class state. One sees that several
but not all off-diagonal elements decay under local dephasing noise,
as determined by $\gamma_{{\rm A}}$. The decoherence timescale for
these terms is thus
\begin{eqnarray}
\tau_{\rm 3-dec,\mathcal{D}}^{\rm W}&=& 2\bigg(\frac{1}{\Gamma_{\rm
1}}\bigg)\ ,
\end{eqnarray}
where the notation
\begin{eqnarray}
\tau_{i-{\rm dec},\mathcal{C}}^{c\left(r\right)},\nonumber\\
\tau_{i-{\rm dis},\mathcal{C}}^{c\left(r\right)},
\end{eqnarray}
represent the decoherence  and disentanglement timescales,
respectively, where $i$ is the number of qubits affected, the
pertinent dephasing noise channel is $\mathcal{C}$ and the state
class is $c$; further special cases may be designated by relative
decay rate, $r$ ( = fast or slow), and is used from here on. The
two-qubit reduced density matrices are then
\begin{eqnarray}
\rho^{{\rm W}}_{{\rm AB}}\left(t\right)=
\left(
\begin{array}{cccc}
 |\bar{a}_{1}|^{2} & 0 & 0 & 0 \\
 0 & |\bar{a}_{2}|^{2} & \bar{a}_{2}\bar{a}_{4}^* \gamma_{\rm A} & 0 \\
 0 & \bar{a}_{4}\bar{a}_{2}^* \gamma_{\rm A} & |\bar{a}_{4}|^{2} & 0 \\
 0 & 0 & 0 & 0
\end{array}\label{WA2QubitAB}
\right),
\\
\rho^{{\rm W}}_{{\rm AC}}\left(t\right)=
\left(
\begin{array}{cccc}
 |\bar{a}_{2}|^{2} & 0 & 0 & 0 \\
 0 & |\bar{a}_{1}|^2 & \bar{a}_{1}\bar{a}_{4}^* \gamma_{\rm A} & 0 \\
 0 & \bar{a}_{4}\bar{a}_{1}^* \gamma_{\rm A} & |\bar{a}_{4}|^2 & 0 \\
 0 & 0 & 0 & 0
 \end{array}\label{WA2QubitAC}
\right),
\\
\rho^{{\rm W}}_{{\rm BC}}\left(t\right)=
\left(
\begin{array}{cccc}
 |\bar{a}_{4}|^{2} & 0 & 0 & 0 \\
 0 & |\bar{a}_{1}|^{2} & \bar{a}_{1}\bar{a}_{2} ^* & 0 \\
 0 & \bar{a}_{2}\bar{a}_{1}^* & |\bar{a}_{2}|^2 & 0 \\
 0 & 0 & 0 & 0
 \end{array}\label{WA2QubitBC}
\right).
\end{eqnarray}

Thus, when part of a triad beginning in a W-class state, qubit pairs
that include the influenced qubit decohere fully, as exhibited by
the presence of the factors $\gamma_{{\rm A}}$ in the expected
off-diagonal terms of the first two of the above density matrices,
whereas subsystem BC maintains its coherence; no such factors appear
in their joint reduced density matrix because in this case the
dephasing channel only acts on qubit A. The two-qubit system
decoherence timescale determined by the factors in the above density
matrix is thus
\begin{eqnarray}
\tau_{\rm 2-dec,\mathcal{D}}^{\rm W}&=& 2\bigg(\frac{1}{\Gamma_{\rm
1}}\bigg).
\end{eqnarray}
Finally, the single-qubit reduced density matrices are
\begin{eqnarray}
\rho^{{\rm W}}_{{\rm A}}\left(t\right)=
\left(
\begin{array}{cc}
 |\bar{a}_{1}|^{2}+|\bar{a}_{2} \left|^2\right. & 0 \\
 0 & |\bar{a}_{4}|^{2}
\end{array}
\right),
\\
\rho^{{\rm W}}_{{\rm B}}\left(t\right)=
\left(
\begin{array}{cc}
  |\bar{a}_{1}|^{2}+|\bar{a}_{4}|^{2} & 0 \\
 0 & |\bar{a}_{2}|^{2}
\end{array}
\right),
\\
\rho^{{\rm W}}_{{\rm C}}\left(t\right)=
\left(
\begin{array}{cc}
 |\bar{a}_{2}|^{2}+|\bar{a}_{4}|^{2} & 0 \\
 0 & |\bar{a}_{1}|^{2} \end{array}
\right),
\end{eqnarray}

\noindent because the individual-qubit matrices are already diagonal
at time $t=0$ and so cannot be dephased further.

Consider now the degrees of entanglement of these two-qubit
subsystems as quantified by the concurrence.
\begin{eqnarray}
C^{2}_{{\rm AB}}&=&
4\left|\bar{a}_{2}\right|^{2}\left|\bar{a}_{4}\right|^{2}\gamma_{{\rm
A}}^{2}
\ ,\\
C^{2}_{{\rm AC}}&=&
4\left|\bar{a}_{1}\right|^{2}\left|\bar{a}_{4}\right|^{2}\gamma_{{\rm
A}}^{2}
\ ,\\
C^{2}_{{\rm BC}}&=&
4\left|\bar{a}_{1}\right|^{2}\left|\bar{a}_{2}\right|^{2} \ .
\end{eqnarray}
The disentanglement timescale  determined by the gamma factor
$\gamma_A$ for the subsystems AB and AC that decohere is thus
\begin{eqnarray}
\tau_{\rm dis,\mathcal{D}}^{\rm W}&=&\bigg(\frac{1}{\Gamma_{\rm
1}}\bigg).
\end{eqnarray}
One sees that disentanglement in subsystems containing a qubit on
which local dephasing noise acts is faster than both the two-qubit
and three-qubit decoherence:
\begin{eqnarray}
\tau_{\rm dis,\mathcal{D}}^{\rm W}&<& \tau_{\rm 3-dec,\mathcal{D}}^{\rm W}\nonumber\ ,\\
\tau_{\rm dis,\mathcal{D}}^{\rm W}&<& \tau_{\rm
2-dec,\mathcal{D}}^{\rm W}\ .
\end{eqnarray}
Neither decoherence nor disentanglement occurs in subsystems not
containing qubit A.


\subsubsection{\label{W2}TWO-QUBIT COLLECTIVE DEPHASING CHANNEL: $\mathcal{E}$}

In the presence of two-qubit collective dephasing noise, acting on
the two-qubit subsystem AB, the time-evolved full three-qubit
density matrix is
\begin{eqnarray}
\rho_{{\rm ABC}}^{{\rm W}}\left(t\right)=
\left(
\begin{scriptsize}
\begin{array}{cccccccc}
 0 & 0 & 0 & 0 & 0 & 0 & 0 & 0 \\
 0 & |\bar{a}_{1}|^2 & \bar{a}_{1}\bar{a}_{2}^* \gamma_{{\rm AB}} & 0 & \bar{a}_{1}\bar{a}_{4}^* \gamma_{{\rm AB}} & 0 & 0 & 0 \\
 0 & \bar{a}_{2}\bar{a}_{1}^* \gamma_{{\rm AB}} & |\bar{a}_{2}|^2 & 0 & \bar{a}_{2}\bar{a}_{4}^*  & 0 & 0 & 0 \\
 0 & 0 & 0 & 0 & 0 & 0 & 0 & 0 \\
 0 & \bar{a}_{4}\bar{a}_{1}^* \gamma_{{\rm AB}} & \bar{a}_{4}\bar{a}_{2}^* & 0 & |\bar{a}_{4}|^2 & 0 & 0 & 0 \\
 0 & 0 & 0 & 0 & 0 & 0 & 0 & 0 \\
 0 & 0 & 0 & 0 & 0 & 0 & 0 & 0 \\
 0 & 0 & 0 & 0 & 0 & 0 & 0 & 0\end{array}
 \end{scriptsize}
\right)\ .
\end{eqnarray}
There is a decay of some but not all off-diagonal density matrix
elements under the local dephasing noise, as determined by
$\gamma_{{\rm AB}}$. A different set of terms now decay compared to
the case of local dephasing noise exhibited in Eq. 70, although in
both cases the terms $\rho_{25}$ and $\rho_{52}$ decay. The
timescale for the decoherence that occurs in this three-qubit
system, determined by the two-qubit gamma factor $\gamma_{\rm AB}$
in the above density matrix, is therefore
\begin{eqnarray}
\tau_{\rm 3-dec,\mathcal{E}}^{\rm W}&=&
2\bigg(\frac{1}{\Gamma_{2}}\bigg).
\end{eqnarray}

The two-qubit reduced density matrices obtained from Eq. 70 are
\begin{eqnarray}
\rho^{{\rm W}}_{{\rm AB}}\left(t\right)=
\left(
\begin{array}{cccc}
 |\bar{a}_{1}|^2 & 0 & 0 & 0 \\
 0 & |\bar{a}_{2}|^2 & \bar{a}_{2}\bar{a}_{4}^* & 0 \\
 0 & \bar{a}_{4}\bar{a}_{2}^* & |\bar{a}_{4}|^2 & 0 \\
 0 & 0 & 0 & 0
 \end{array}
\right)\ ,
\\
\rho^{{\rm W}}_{{\rm AC}}\left(t\right)=
\left(
\begin{array}{cccc}
 |\bar{a}_{2}|^2 & 0 & 0 & 0 \\
 0 & |\bar{a}_{1}|^{2} & \bar{a}_{1}\bar{a}_{4}^* \gamma _{{\rm AB}} & 0 \\
 0 & \bar{a}_{4}\bar{a}_{1}^* \gamma _{{\rm AB}} & |\bar{a}_{4}|^2 & 0 \\
 0 & 0 & 0 & 0
\end{array}
\right)\ ,
\\
\rho^{{\rm W}}_{{\rm BC}}\left(t\right)=
\left(
\begin{array}{cccc}
 |\bar{a}_{4}|^2 & 0 & 0 & 0 \\
 0 & |\bar{a}_{1}|^2 & \bar{a}_{1}\bar{a}_{2}^* \gamma_{{\rm AB}} & 0 \\
 0 & \bar{a}_{2}\bar{a}_{1}^* \gamma_{{\rm AB}} & |\bar{a}_{2}|^2 & 0 \\
 0 & 0 & 0 & 0
 \end{array}
\right)\ .
\end{eqnarray}
It is significant that the off-diagonal elements of $\rho_{{\rm
W}}^{{\rm AB}}$ \emph{do not} decay, whereas the those of
$\rho_{{\rm W}}^{{\rm AC}}$ and $\rho_{{\rm W}}^{{\rm BC}}$ do. By
comparing these expressions with those of Eqs. 73-75, we also see
that collective noise acting on subsystems including only one
affected qubit has an effect analogous to that of the local noise
environment on that channel. This behavior also appears below in the
case of total-system collective dephasing of initially W-class
states, where a group of qubits subjected to the fully collective
noise also do not lose decoherence as a result of that collective
noise. The timescale of two-qubit system decoherence determined by
the factors in the above density matrices, when it occurs, is thus
\begin{eqnarray}
\tau_{\rm 2-dec,\mathcal{E}}^{\rm
W}=2\bigg(\frac{1}{\Gamma_{2}}\bigg).
\end{eqnarray}
Again single-qubit reduced density matrices are diagonal from the
outset, {\it cf.} Eqs. 77-79, and so cannot decohere further.

The degrees of entanglement of two-qubit subsystems are
\begin{eqnarray}
C^{2}_{{\rm AB}}&=&
4\left|\bar{a}_{2}\right|^{2}\left|\bar{a}_{4}\right|^{2}\ ,\\
C^{2}_{{\rm AC}}&=&
4\left|\bar{a}_{1}\right|^{2}\left|\bar{a}_{4}\right|^{2}\gamma_{{\rm
AB}}^{2}\ ,\\
C^{2}_{{\rm BC}}&=&
4\left|\bar{a}_{1}\right|^{2}\left|\bar{a}_{2}\right|^{2}\gamma_{{\rm
AB}}^{2}\ .
\end{eqnarray}
Comparing these expressions with the local-dephasing noise scenario
expressions in Eqs. 80-82, one again sees similar behavior when only
one qubit of a subsystem is subject to collective noise. The
timescale of disentanglement determined by the factors in the above
density matrix, when it occurs, is thus
\begin{eqnarray}
\tau_{\rm dis,\mathcal{E}}^{\rm W}=\bigg(\frac{1}{\Gamma_{\rm
2}}\bigg)\ .
\end{eqnarray}
In this case, one again sees that disentanglement, which occurs only
when decoherence occurs, is always faster than three-qubit and
two-qubit decoherence in all instances in which these effects occur:
\begin{eqnarray}
\tau_{\rm dis,\mathcal{E}}^{\rm W} < \tau_{\rm 3-dec,\mathcal{E}}^{\rm W}\nonumber\ ,\\
\tau_{\rm dis,\mathcal{E}}^{\rm W} < \tau_{\rm
2-dec,\mathcal{E}}^{\rm W} \ .
\end{eqnarray}

\subsubsection{\label{W3}THREE-QUBIT COLLECTIVE DEPHASING CHANNEL:
$\mathcal{F}$}

The time-evolved three-qubit density matrix, however, shows no
effect from the three-qubit collective dephasing channel:
\begin{widetext}
\begin{eqnarray}
\rho_{{\rm ABC}}^{{\rm W}}\left(t\right)= \left(
\begin{array}{cccccccc}
 0 & 0 & 0 & 0 & 0 & 0 & 0 & 0 \\
 0 & |\bar{a}_{1}|^2 & \bar{a}_{1}\bar{a}_{2}^* & 0 & \bar{a}_{1}\bar{a}_{4}^* & 0 & 0 & 0 \\
 0 & \bar{a}_{2}\bar{a}_{1}^* & |\bar{a}_{2}|^2 & 0 & \bar{a}_{2}\bar{a}_{4}^* & 0 & 0 & 0 \\
 0 & 0 & 0 & 0 & 0 & 0 & 0 & 0 \\
 0 & \bar{a}_{4}\bar{a}_{1}^* & \bar{a}_{4}\bar{a}_{2}^* & 0 & |\bar{a}_{4}|^2 & 0 & 0 & 0 \\
 0 & 0 & 0 & 0 & 0 & 0 & 0 & 0 \\
 0 & 0 & 0 & 0 & 0 & 0 & 0 & 0 \\
 0 & 0 & 0 & 0 & 0 & 0 & 0 & 0\end{array}
\right)\ .
\end{eqnarray}
\end{widetext}
Hence, the W-class states can be identified as a class that is
entirely robust under fully collective noise, in the sense being
``decoherence-free,'' which can be attributed to the permutation
symmetry introduced with the initial state $|W^g\rangle$,
analogously to what occurs in two-qubit systems under collective
dephasing ({\it cf.} \cite{Kwiat}). For the GHZ-class states,
however, one finds decoherence does occur, as shown in Section C,
below.


\subsubsection{\label{WthreeLocal} THREE-QUBIT MULTI-LOCAL DEPHASING:
$\mathcal{D}^{\rm A}\mathcal{D}^{\rm B}\mathcal{D}^{\rm C}$}

Under multi-local dephasing noise, wherein each qubit is
individually subject to the same sort of noise, one finds the
time-evolved density matrix arising from initially W-class states to
be
\begin{eqnarray}
\rho_{{\rm ABC}}^{{\rm W}}\left(t\right)= \left(
\begin{array}{cccccccc}
 0 & 0 & 0 & 0 & 0 & 0 & 0 & 0 \\
 0 & |\bar{a}_{1}|^2 & \bar{a}_{1}\bar{a}_{2}^* \gamma_{{\rm B}} \gamma_{{\rm C}} & 0 & \bar{a}_{1}\bar{a}_{4}^* \gamma_{{\rm A}} \gamma_{{\rm C}} & 0 & 0 & 0 \\
 0 & \bar{a}_{2}\bar{a}_{1}^* \gamma_{{\rm B}} \gamma_{{\rm C}} & |\bar{a}_{2}|^2 & 0 & \bar{a}_{2}\bar{a}_{4}^* \gamma_{{\rm A}} \gamma_{{\rm B}} & 0 & 0 & 0 \\
 0 & 0 & 0 & 0 & 0 & 0 & 0 & 0 \\
 0 & \bar{a}_{4}\bar{a}_{1}^* \gamma_{{\rm A}} \gamma_{{\rm C}} & \bar{a}_{4}\bar{a}_{2}^* \gamma_{{\rm A}} \gamma_{{\rm B}} & 0 & |\bar{a}_{4}|^2 & 0 & 0 & 0 \\
 0 & 0 & 0 & 0 & 0 & 0 & 0 & 0 \\
 0 & 0 & 0 & 0 & 0 & 0 & 0 & 0 \\
 0 & 0 & 0 & 0 & 0 & 0 & 0 & 0\end{array}
\right)\ ,
\end{eqnarray}
wherein all off-diagonal elements decay according to a product of
two differing $\gamma$ factors. The three-qubit system decoherence
timescale determined by those factors (recalling that
$\gamma_A=\gamma_B=\gamma_C$) in the above density matrices is thus
\begin{eqnarray}
\tau_{{\rm 3-dec},\mathcal{D}^{\rm A}\mathcal{D}^{\rm
B}\mathcal{D}^{\rm C}}^{\rm W}&=&\bigg(\frac{1}{\Gamma_{1}}\bigg)\ .
\end{eqnarray}

The two-qubit reduced density matrices are
\begin{eqnarray}
\rho^{{\rm W}}_{{\rm AB}}\left(t\right)=
\left(
\begin{array}{cccc}
 |\bar{a}_{1}|^2 & 0 & 0 & 0 \\
 0 & |\bar{a}_{2}|^2 & \bar{a}_{2}\bar{a}_{4}^* \gamma_{{\rm A}} \gamma_{{\rm B}} & 0 \\
 0 & \bar{a}_{4}\bar{a}_{2}^* \gamma_{{\rm A}} \gamma_{{\rm B}} & |\bar{a}_{4}|^2 & 0 \\
 0 & 0 & 0 & 0
 \end{array}
\right)\ ,
\\
\rho^{{\rm W}}_{{\rm AC}}\left(t\right)=
\left(
\begin{array}{cccc}
 |\bar{a}_{2}|^2 & 0 & 0 & 0 \\
 0 & |\bar{a}_{1}|^2 & \bar{a}_{1}\bar{a}_{4}^* \gamma_{{\rm A}} \gamma_{{\rm C}} & 0 \\
 0 & \bar{a}_{4}\bar{a}_{1}^* \gamma_{{\rm A}} \gamma_{{\rm C}} & |\bar{a}_{4}|^2 & 0 \\
 0 & 0 & 0 & 0
\end{array}
\right)\ ,
\\
\rho^{{\rm W}}_{{\rm BC}}\left(t\right)=
\left(
\begin{array}{cccc}
 |\bar{a}_{4}|^2 & 0 & 0 & 0 \\
 0 & |\bar{a}_{1}|^2 & \bar{a}_{1}\bar{a}_{2}^* \gamma_{{\rm B}} \gamma_{{\rm C}} & 0 \\
 0 & \bar{a}_{2}\bar{a}_{1}^* \gamma_{{\rm B}} \gamma_{{\rm C}} & |\bar{a}_{2}|^2 & 0 \\
 0 & 0 & 0 & 0
 \end{array}
\right)\ ,
\end{eqnarray}
similar in form to those in Eqs. 73-75 but having a $\gamma_X$
factor in off-diagonal elements for each corresponding local (to
qubit $X$) noise source. One sees that all off-diagonal elements
decay, as determined by two pertinent $\gamma$ factors.  The
two-qubit system decoherence timescale determined by the factors in
the above density matrices is similarly thus
\begin{eqnarray}
\tau_{\rm 2-dec,\mathcal{D}^{\rm A}\mathcal{D}^{\rm
B}\mathcal{D}^{\rm C}}^{\rm W}&=&\bigg(\frac{1}{\Gamma_{1}}\bigg)\ .
\end{eqnarray}
The single-qubit reduced density matrices are again diagonal, as
shown in Eqs. 77-79, and so are not able to dephase further.

The degrees of entanglement of the two-qubit subsystems given by
concurrence are
\begin{eqnarray}
C^{2}_{{\rm AB}}&=&
4\left|\bar{a}_{2}\right|^{2}\left|\bar{a}_{4}\right|^{2}\gamma_{{\rm
A}}^{2}\gamma_{{\rm B}}^{2}
\ ,\\
C^{2}_{{\rm AC}}&=&
4\left|\bar{a}_{1}\right|^{2}\left|\bar{a}_{4}\right|^{2}\gamma_{{\rm
A}}^{2}\gamma_{{\rm C}}^{2}
\ ,\\
C^{2}_{{\rm BC}}&=&
4\left|\bar{a}_{1}\right|^{2}\left|\bar{a}_{2}\right|^{2}\gamma_{{\rm
B}}^{2}\gamma_{{\rm C}}^{2} \ ,
\end{eqnarray}
with the particular $\gamma$ changes reflecting the dephasing
channel. The disentanglement timescale is thus
\begin{eqnarray}
\tau_{{\rm 2-dis},\mathcal{D}^{\rm A}\mathcal{D}^{\rm
B}\mathcal{D}^{\rm C}}^{\rm W}&=&{1\over
2}\bigg(\frac{1}{\Gamma_{\rm 1}}\bigg)\ .
\end{eqnarray}
We thus find again that disentanglement is always faster than
three-qubit decoherence and two-qubit decoherence
\begin{eqnarray}
\tau_{{\rm dis},\mathcal{D}^{\rm A}\mathcal{D}^{\rm
B}\mathcal{D}^{\rm C}}^{\rm W}&<& \tau_{\rm 3-dec,\mathcal{D}^{\rm
A}\mathcal{D}^{\rm
B}\mathcal{E}^{\rm C}}^{\rm W}\nonumber\ ,\\
\tau_{{\rm dis},\mathcal{D}^{\rm A}\mathcal{D}^{\rm
B}\mathcal{D}^{\rm C}}^{\rm W}&<& \tau_{\rm 2-dec,\mathcal{D}^{\rm
A}\mathcal{D}^{\rm B}\mathcal{D}^{\rm C}}^{\rm W}\ ,
\end{eqnarray}
for multi-local dephasing noise.
\subsubsection{\label{WLAGAB}ONE-QUBIT LOCAL, TWO-QUBIT COLLECTIVE
DEPHASING $\mathcal{D}\mathcal{E}$}

For three-qubit systems, the possibility of combinations of local
and subsystem-collective dephasing noise exists, which we now
consider. Under such noise,  where qubit A, for example, is affected
by local noise and the remaining subsystem BC is subject to
collective noise ({\it i.e.} noise collective at the scale of BC),
the time-evolved state is
\begin{widetext}
\begin{eqnarray}
\rho_{{\rm ABC}}^{{\rm W}}\left(t\right)= \left(
\begin{array}{cccccccc}
 0 & 0 & 0 & 0 & 0 & 0 & 0 & 0 \\
 0 & |\bar{a}_{1}|^2 & \bar{a}_{1}\bar{c}_{2} & 0 & \bar{a}_{1}\bar{c}_{4}\gamma_{{\rm A}}\gamma_{{\rm BC}} & 0 & 0 & 0 \\
 0 & \bar{a}_{2}\bar{c}_{1} & |\bar{a}_{2}|^2 & 0 & \bar{a}_{2}\bar{c}_{4}\gamma_{{\rm A}}\gamma_{{\rm BC}} & 0 & 0 & 0 \\
 0 & 0 & 0 & 0 & 0 & 0 & 0 & 0 \\
 0 & \bar{a}_{4}\bar{c}_{1}\gamma_{{\rm A}}\gamma_{{\rm BC}} & \bar{a}_{4}\bar{c}_{2}\gamma_{{\rm A}}\gamma_{{\rm BC}} & 0 & |\bar{a}_{4}|^2 & 0 & 0 & 0 \\
 0 & 0 & 0 & 0 & 0 & 0 & 0 & 0 \\
 0 & 0 & 0 & 0 & 0 & 0 & 0 & 0 \\
 0 & 0 & 0 & 0 & 0 & 0 & 0 & 0
\end{array}
\right)\ .
\end{eqnarray}
\end{widetext}
The effective three-qubit system decoherence timescale determined by
the factors in the above density matrices is thus
\begin{eqnarray}
\tau_{{\rm 3-dec},\mathcal{D}\mathcal{E}}^{\rm W}&=&
2\bigg(\frac{1}{\Gamma_{1}}\bigg) +
2\bigg(\frac{1}{\Gamma_{2}}\bigg)\ .
\end{eqnarray}
In this case, one finds slightly different decay behavior than in
cases above: the total decay rate is composed of a product of
$\gamma$ factors. In this case and in subsequent similar cases, we
will assume, for clarity of exposition, that either one of the two
$\gamma$ factors, one corresponding to local noise and the other to
collective noise, is much larger or that the two are of comparable
strength.

The two-qubit reduced density matrices in this case are
\begin{eqnarray}
\rho^{{\rm W}}_{{\rm AB}}\left(t\right)=
\left(
\begin{array}{cccc}
 |\bar{a}_{1}|^2 & 0 & 0 & 0 \\
 0 & |\bar{a}_{2}|^2 & \bar{a}_{2}\bar{c}_{4}\gamma_{{\rm A}}\gamma_{{\rm BC}} & 0 \\
 0 & \bar{a}_{4}\bar{c}_{2}\gamma_{{\rm A}}\gamma_{{\rm BC}} & |\bar{a}_{4}|^2 & 0 \\
 0 & 0 & 0 & 0
\end{array}
\right),\\
\nonumber\\
\rho^{{\rm W}}_{{\rm AC}}\left(t\right)=
\left(
\begin{array}{cccc}
 |\bar{a}_{2}|^2 & 0 & 0 & 0 \\
 0 & |\bar{a}_{1}|^2 & \bar{a}_{1}\bar{c}_{4}\gamma_{{\rm A}}\gamma_{{\rm BC}} & 0 \\
 0 & \bar{a}_{4}\bar{c}_{1}\gamma_{{\rm A}}\gamma_{{\rm BC}} & |\bar{a}_{4}|^2 & 0 \\
 0 & 0 & 0 & 0
\end{array}
\right),\\
\nonumber\\
\rho^{{\rm W}}_{{\rm BC}}\left(t\right)=
\left(
\begin{array}{cccc}
 |\bar{a}_{4}|^2 & 0 & 0 & 0 \\
 0 & |\bar{a}_{1}|^2 & \bar{a}_{1}\bar{c}_{2} & 0 \\
 0 & \bar{a}_{2}\bar{c}_{1} & |\bar{a}_{2}|^2 & 0 \\
 0 & 0 & 0 & 0
\end{array}
\right).
\end{eqnarray}
The off-diagonal elements also decay due to the appearance of a
combination of the two pertinent $\gamma$ factors in the first two
subsystems above. Because the two qubits of subsystem BC are
subjected to collective dephasing noise \emph{and} because W-class
states are symmetrized, this subsystem does not decohere, differing
from the case of two-qubit subsystem-collective dephasing only and
the three-qubit collective dephasing cases. The two-qubit system
decoherence timescale, in subsystems AB and AC, determined by these
factors is thus
\begin{eqnarray}
\tau_{{\rm 2-dec},\mathcal{D}\mathcal{E}}^{\rm W}&& =
2\bigg(\frac{1}{\Gamma_{1}}\bigg) +
2\bigg(\frac{1}{\Gamma_{2}}\bigg) \ ,
\end{eqnarray}
the same combination of rates as for three-qubit decoherence.

Recall again that the single-qubit reduced density matrices, given
by Eqs. 77-79, are fully dephased from the outset.

The entanglement of subsystems given by the concurrences
\begin{eqnarray}
C^{2}_{{\rm AB}}&=&
4\left|\bar{a}_{2}\right|^{2}\left|\bar{a}_{4}\right|^{2}\gamma_{{\rm
A}}^{2}\gamma_{{\rm BC}}^{2}
\ ,\\
C^{2}_{{\rm AC}}&=&
4\left|\bar{a}_{1}\right|^{2}\left|\bar{a}_{4}\right|^{2}\gamma_{{\rm
A}}^{2}\gamma_{{\rm BC}}^{2}
\ ,\\
C^{2}_{{\rm BC}}&=&
4\left|\bar{a}_{1}\right|^{2}\left|\bar{a}_{2}\right|^{2}\ ,
\end{eqnarray}
has behavior determined by the factors in the above density
matrices, so that for subsystems AB and AC
\begin{eqnarray}
\tau_{{\rm dis},\mathcal{D}\mathcal{E}}^{\rm W}&& =
\bigg(\frac{1}{\Gamma_{1}}\bigg) + \bigg(\frac{1}{\Gamma_{2}}\bigg)\
.
\end{eqnarray}
Disentanglement is therefore always faster than two-qubit and
three-qubit decoherence under this noise,
\begin{eqnarray}
\tau_{{\rm dis},\mathcal{D}\mathcal{E}}^{\rm W} <
\tau_{{\rm 3-dec},\mathcal{D}\mathcal{E}}^{\rm W}\nonumber\ ,\\
\tau_{{\rm dis},\mathcal{D}\mathcal{E}}^{\rm W} < \tau_{{\rm
2-dec},\mathcal{D}\mathcal{E}}^{\rm W},
\end{eqnarray}
when  it occurs.


\subsubsection{\label{Wresults}DISCUSSION}

From the above results for the W class of states, we have shown that
the disentanglement timescale is always faster than or equal to the
decoherence rate as described by the off-diagonal terms in all five
dephasing-noise environments, namely, one-qubit local
($\mathcal{D}$), two-qubit collective ($\mathcal{E}$), three-qubit
collective ($\mathcal{F}$), three-qubit multi-local
($\mathcal{DDD}$), and one-qubit local plus two-qubit collective
($\mathcal{D}\mathcal{E}$) environments. Thus, the decoherence rate
bounds the disentanglement rate from below for systems beginning out
in W-class states.

\subsection{\label{behaviorGHZ}BEHAVIOR OF GHZ-CLASS STATES}

The analysis of the GHZ-class of states will proceed in a similar
fashion as that of the W-class but can be explicated more briefly.
The initial density matrix of the GHZ-class state of Eq. 46 is
\begin{eqnarray}
\rho^{{\rm GHZ}}_{{\rm ABC}}\left(0\right)=P(\ket{\rm GHZ^g}) \equiv
\left(
\begin{array}{cccccccc}
 |\bar{a}_{0}|^{2} & 0 & 0 & 0 & 0 & 0 & 0 & \bar{a}_{0}\bar{a}_{7}^* \\
 0 & 0 & 0 & 0 & 0 & 0 & 0 & 0 \\
 0 & 0 & 0 & 0 & 0 & 0 & 0 & 0 \\
 0 & 0 & 0 & 0 & 0 & 0 & 0 & 0 \\
 0 & 0 & 0 & 0 & 0 & 0 & 0 & 0 \\
 0 & 0 & 0 & 0 & 0 & 0 & 0 & 0 \\
 0 & 0 & 0 & 0 & 0 & 0 & 0 & 0 \\
 \bar{a}_{7}\bar{a}_{0}^* & 0 & 0 & 0 & 0 & 0 & 0 & |\bar{a}_{0}|^{2}
\end{array}
\right).
\end{eqnarray}
\subsubsection{\label{ghzDephasingChannels}DEPHASING CHANNELS}

The effect of all dephasing noise types upon the GHZ-class density
matrix is similar, differing only by the representative decay factor
$\tilde\gamma$ of the decay channel, in the five previously
considered noise-scenarios:
\begin{eqnarray}
\mathcal{D}^{\rm A}:&&\tilde{\gamma}=\gamma_{\rm A}\ \ \ \ {\rm 1-qubit\ local} \nonumber ,\\
\mathcal{E}^{\rm AB}:&&\tilde{\gamma}=\gamma_{\rm AB}^{4}\ \ \ \ {\rm 2-qubit\ local}\nonumber ,\\
\mathcal{F}:&&\tilde{\gamma}=\gamma_{\rm ABC}^{4}\ \ \ \ {\rm 3-qubit\ collective}\nonumber ,\\
\mathcal{D}^{\rm A}\mathcal{D}^{\rm B}\mathcal{D}^{\rm C}:&&\tilde{\gamma}=\gamma_{\rm A}\gamma_{\rm B}\gamma_{\rm C}\ \ \ \ {\rm multi-local}\nonumber ,\\
\mathcal{D}^{\rm A}\mathcal{E}^{\rm BC}:&&\tilde{\gamma}=\gamma_{\rm
A}\gamma_{\rm BC}^{4}\ \ \ \ {\rm 1-qubit\ local\ and\ two-qubit\
collective} \ ,
\end{eqnarray}
which appear in all of the off-diagonal elements in the density
matrices in the following time-evolved density matrix, so that
\begin{eqnarray}
\rho^{{\rm GHZ}}_{{\rm ABC}}\left(t\right)=
\left(
\begin{array}{cccccccc}
 |\bar{a}_{0}|^2 & 0 & 0 & 0 & 0 & 0 & 0 & \bar{a}_{0}\bar{a}_{7}^{*} \tilde{\gamma} \\
 0 & 0 & 0 & 0 & 0 & 0 & 0 & 0 \\
 0 & 0 & 0 & 0 & 0 & 0 & 0 & 0 \\
 0 & 0 & 0 & 0 & 0 & 0 & 0 & 0 \\
 0 & 0 & 0 & 0 & 0 & 0 & 0 & 0 \\
 0 & 0 & 0 & 0 & 0 & 0 & 0 & 0 \\
 0 & 0 & 0 & 0 & 0 & 0 & 0 & 0 \\
 \bar{a}_{7}\bar{a}_{0}^{*} \tilde{\gamma} & 0 & 0 & 0 & 0 & 0 & 0 & |\bar{a}_{7}|^2
\end{array}
\right)\ ,
\end{eqnarray}
which decays to a diagonal matrix as determined by the particular
form of $\tilde{\gamma}$ under the influence of the noise models
introduced in the previous section. The three-qubit-system
decoherence timescales are thus
\begin{eqnarray}
\tau_{{\rm 3-dec},\mathcal{D}^{\rm A}}^{\rm GHZ}
&=& 2\bigg(\frac{1}{\Gamma_{1}}\bigg) \ ,\\
\tau_{{\rm 3-dec},\mathcal{E}^{\rm AB}}^{\rm GHZ}
&=& {1\over 2}\bigg(\frac{1}{\Gamma_{2}}\bigg) \ ,\\
\tau_{{\rm 3-dec},\mathcal{F}^{\rm ABC}}^{\rm GHZ}
&=& {1\over 2}\bigg(\frac{1}{\Gamma_{3}}\bigg) \ ,\\
\tau_{{\rm 3-dec},\mathcal{D}^{\rm A}\mathcal{D}^{\rm
B}\mathcal{D}^{\rm C}}^{\rm GHZ}
&=& \frac{2}{3}\bigg({1\over\Gamma_{1}}\bigg) \ ,\\
\tau_{{\rm 3-dec},\mathcal{D}^{\rm A}\mathcal{E}^{\rm BC}}^{\rm GHZ}
&=& 2\bigg(\frac{1}{\Gamma_{1}}\bigg) + {1\over
2}\bigg(\frac{1}{\Gamma_{2}}\bigg) \ .
\end{eqnarray}
The two-qubit reduced density matrices are, unlike the case of the W
class, seen to be diagonal from the outset and so cannot decohere or
dephase further:
\begin{eqnarray}
\rho^{{\rm GHZ}}_{{\rm AB}}\left(t\right)=
\left(
\begin{array}{cccc}
 |\bar{a}_{0}|^2 & 0 & 0 & 0 \\
 0 & 0 & 0 & 0 \\
 0 & 0 & 0 & 0 \\
 0 & 0 & 0 & |\bar{a}_{7}|^2
\end{array}
\right)\ ,
\\
\rho^{{\rm GHZ}}_{{\rm AC}}\left(t\right)=
\left(
\begin{array}{cccc}
 |\bar{a}_{0} |^{2} & 0 & 0 & 0 \\
 0 & 0 & 0 & 0 \\
 0 & 0 & 0 & 0 \\
 0 & 0 & 0 & |\bar{a}_{7}|^2
 \end{array}
\right)\ ,
\\
\rho^{{\rm GHZ}}_{{\rm BC}}\left(t\right)=
\left(
\begin{array}{cccc}
 |\bar{a}_{0} |^{2} & 0 & 0 & 0 \\
 0 & 0 & 0 & 0 \\
 0 & 0 & 0 & 0 \\
 0 & 0 & 0 & |\bar{a}_{7} |^{2}
 \end{array}
\right)\ .
\end{eqnarray}
The single-qubit reduced density matrices are similarly always
diagonal:
\begin{eqnarray}
\rho^{{\rm GHZ}}_{{\rm A}}\left(t\right)=
\left(
\begin{array}{cc}
 |\bar{a}_{0} |^{2} & 0 \\
 0 & |\bar{a}_{7} |^{2}
\end{array}
\right)\ ,
\\
\rho^{{\rm GHZ}}_{{\rm B}}\left(t\right)=
\left(
\begin{array}{cc}
 |\bar{a}_{0} |^{2} & 0 \\
 0 & |\bar{a}_{7} |^{2}
\end{array}
\right)\ ,
\\
\rho^{{\rm GHZ}}_{{\rm C}}\left(t\right)=
\left(
\begin{array}{cc}
 |\bar{a}_{0} |^{2} & 0 \\
 0 & |\bar{a}_{7}|^{2}
\end{array}
\right)\ .
\end{eqnarray}
For two-qubit-subsystem entanglement, one finds
\begin{eqnarray}
C^{2}_{{\rm AB}}&=& 0
\ ,\\
C^{2}_{{\rm AC}}&=& 0
\ ,\\
C^{2}_{{\rm BC}}&=& 0 \ .
\end{eqnarray}

The reduced states of all proper subsystems in states initially of
the GHZ-class are neither coherent nor entangled \emph{at any time},
when subject to all the above forms of dephasing noise; all
entanglement appears only at the triple-qubit level, for which there
is no analytic mixed-state entanglement measure of which we are
aware. Nonetheless, one sees that, at the two-qubit level, the
behavior of this class of states still formally agrees with our
general conclusion regarding the W-class, with the difference that
the two-qubit (lack of) decoherence trivially bounds the trivial
(lack of) bipartite entanglement. Unfortunately, as in the case of
the W-class states but more significantly in this case, the lack of
a well defined analytic mixed-state three-qubit entanglement measure
precludes one from performing a detailed comparison of three-qubit
disentanglement versus decoherence at the three-qubit level, which
would be useful.

\section{\label{conclusion}CONCLUSIONS}
We have shown for the fundamental three-qubit entanglement classes
that there are different timescales for two-qubit disentanglement
and decoherence
under dephasing noise. In particular, we have shown for three qubits
initially in an entangled state subject to an array of combinations
of dephasing-noise types, disentanglement, when it occurs, does so
on a timescale shorter than or equal to that of decoherence. In the
course of obtaining these results, we determined the precise
dephasing and bipartite disentanglement behavior for each of the
various pertinent classes of two-qubit and three-qubit states under
all dephasing noise scenarios available at their respective scales,
and have clearly identified dephasing-free and disentanglement-free
classes. The coherence and bipartite entanglement behavior for these
classes has thereby been provided for future investigators, who can
now select and/or engineer states for practical applications, such
as quantum metrology, communication and information processing,
according to the coherence and entanglement characteristics these
tasks require.

\end{document}